# Magnetic interactions based on proton orbital motion in $CH_3NH_3PbI_3$ and $CH_3NH_3PbBr_3$

Lei Meng,[a,†] Miao Miao Zhao,[a,†] Yi Yang Xu,[a,†] Chu Xin Peng,[a,†] Yang Yang,[a] Tian Tian Xing,[a] Peng Ren[a,*] and Fei Yen[a,*]

a. Department State Key Laboratory on Tunable Laser Technology, Ministry of Industry and Information Technology Key Laboratory of Micro-Nano Optoelectronic Information System and the School of Science, Harbin Institute of Technology, Shenzhen. University Town, Shenzhen, Guangdong 518055, P.R. China.
*Corresponding authors: fyen@hit.edu.cn (Tel: +86-1343-058-9183), renpeng@hit.edu.cn
†These authors contributed equally.

**Abstract:** The microscopic origin of the remarkable optoelectronic properties of one of the most studied contemporary materials remains unclear. Here, we identify the existence of magnetic interactions between intermolecular proton orbitals in $CH_3NH_3PbI_3$ and $CH_3NH_3PbBr_3$. In particular, a unique sharp drop and a pronounced step-up discontinuity in the magnetic susceptibility at the tetragonal-to-cubic phase transitions are identified in $CH_3NH_3PbI_3$ and $CH_3NH_3PbBr_3$, respectively. The magnetic interactions in the orthorhombic and tetragonal phases are dependent on thermal history and lattice orientation while nearly independent of the applied external magnetic field. In $CH_3NH_3PbBr_3$, the $CH_3$ and $NH_3^+$ components reorient in an uncorrelated fashion resulting the cubic phase to also exhibit magnetic anisotropy. Our findings provide a potential link connecting the highly light-absorbing $CH_3NH_3^+$ and the exceptional properties of the charge carriers of the inorganic framework in hybrid perovskite solar cells.

Methylammonium lead halides remain an extremely captivating family of materials because of their now regularly over-20% photovoltaic power conversion efficiencies [1,2] amongst many other exceptional properties and functionalities [3-7]. However, the fundamental mechanism(s) underpinning its many optoelectronic properties remains unclear and continue to be a subject of intense debate [8]. As one of the most intensely studied compounds, practically all types of experiments have already been undertaken, except one: temperature-dependent magnetic susceptibility measurements of single crystals. The reason for this is because these salts are diamagnetic, like water, wood and most plastics, all of the electrons are paired up so the magnetic susceptibility is independent of temperature according to the theory of Langevin diamagnetism. Now recently, we identified that apart from electrons, neighboring protons ($H^+$) experiencing orbital motion (such as $NH_4^+$ ions continually reorienting by 90° or 120°) become magnetically resonant with each other if their orbitals are the same [9-12]. At room temperature neighboring ions have a large enough selection of orbitals to avoid mutual resonance, but as the temperature is lowered, subsets of proton orbitals become energetically inaccessible so the lattice becomes highly resonant and unstable. As a means to avoid structural collapse the ions become ordered so resonant forces can cancel each other out. The ordering process of the proton orbitals shows up in the magnetic susceptibility so temperature and external magnetic field scans revealed a great deal of information on the lattice dynamics of many molecular solids [9-13]. When the reorienting ion is polar such as $CH_3NH_3^+$, two types of resonant forces arise in the ordered ground state because the $CH_3$ and $NH_3^+$ reorient at different rates: the C–H and N–H distances are different and they also form distinct hydrogen bonds. From such, two non-equivalent resonant frameworks emerge and the lattice distorts into a polar phase at the order-disorder phase transitions to yield ferroelectricity such as the β and γ phases of $CH_3NH_3Cl$ and $CH_3NH_3Br$ identified recently [13]. This is essentially the molecular analogue of when two types of electron spins become long-range ordered and break spatial-inversion symmetry [14]. We believe this mechanism underpins the ferroelectric origin in the orthorhombic and tetragonal phases of $CH_3NH_3PbBr_3$ (MAPB) [15] since nuclear magnetic resonance (NMR) measurements showed unambiguously that the $CH_3$ and $NH_3^+$ components of a fraction of $CH_3NH_3^+$ reorient uncorrelated [16]. With $CH_3NH_3PbI_3$ (MAPI), the story is different in that the $CH_3$ and $NH_3^+$ groups continue to reorient in a correlated fashion all the way down to the low temperature phase [16]. Nevertheless, there are still resonant effects from the $CH_3$ and $NH_3^+$ groups except that their distortions end up nearly cancelling each other out to render MAPI weakly ferrielectric to antiferroelectric. A presence of ferroelectricity is strongly believed to reside in the methylammonium lead halides because a polar crystal structure gives rise to second-order optical nonlinearity and a linear electro-optic effect [17] which explains why there are numerous reports on the existence [18-21] and nonexistence [22-24] of ferroelectricity in MAPI. For these reasons, it is of utmost importance to study the magnetic properties of MAPI and MAPB to verify whether orbital-orbital interactions are present in the system. Observation of magnetoelectric coupling in MAPI and MAPB will offer researchers the possible scenario that proton orbital-orbital magnetic resonant interactions play a prominent role in the remarkable optoelectronic properties of perovskite solar cells.

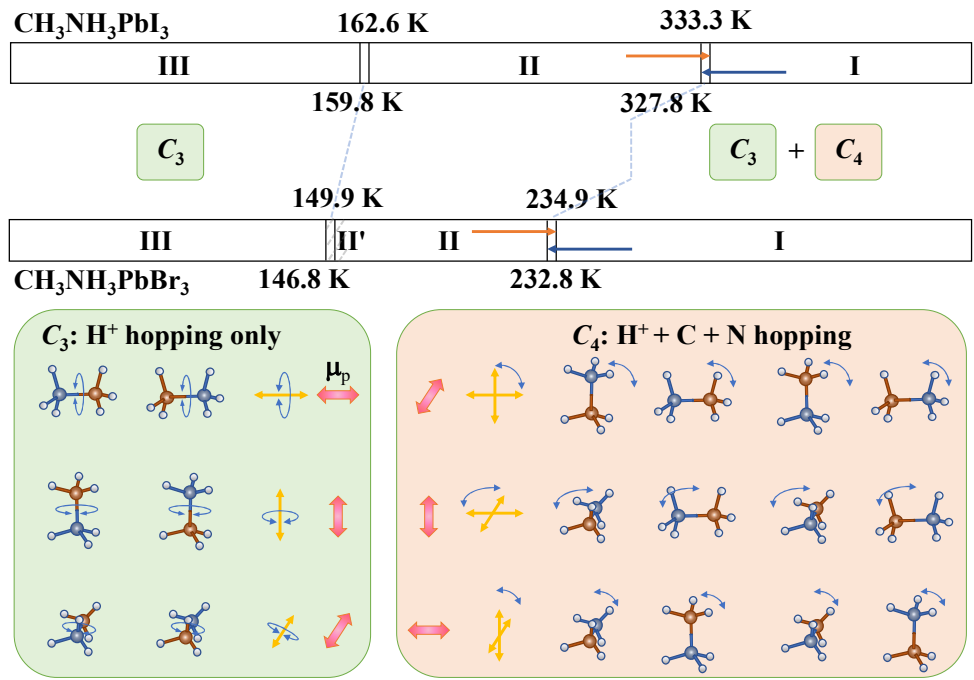

**Fig. 1.** Structural phase transitions and $CH_3NH_3^+$ dynamics of $CH_3NH_3PbI_3$ and $CH_3NH_3PbBr_3$. Top Panel: Stability regions of the cubic (I), tetragonal (II and II') and orthorhombic (III) phases. Bottom panels show the $C_3$ and $C_4$ types of motions of the $CH_3NH_3^+$ ions with respect to the principal axes (yellow arrows) of the crystal lattice and their associated magnetic moments $\mu_p$ (pink arrows).

Figure 1 shows the temperature regions of the polymorphs of MAPI and MAPB at ambient pressure. Starting at high temperatures the lattice structure is cubic (Phase I); with cooling, the systems first phase transition to become tetragonal (Phase II) and then orthorhombic (Phase III) [25]. MAPB is slightly more complex as it possesses a second tetragonal phase (Phase II'). The space groups of these phases have been claimed to be centrosymmetric and non-centrosymmetric and continues to be a topic under debate [18-21,26-28]. What is important to us are the fluxional aspects of the $CH_3NH_3^+$ ions. According to neutron scattering [29] and NMR experiments [16,30], the C–N bonds in Phases I and II experience three-fold reorientations ($C_3$) about its own axis as well as four-fold reorientations about the *a*, *b* and *c* principal axes ($C_4$) [16,29,30] as illustrated in the bottom panels of Fig. 1. In Phase III, the more energetic $C_4$ reorientations seize while the $C_3$ motions continue. For the case of MAPI, the $CH_3$ and $NH_3^+$ components reorient in a correlated fashion (locked with each other) in all three phases. In contrast for the case of MAPB, a fraction of the $CH_3$ and $NH_3^+$ in Phase III reorient uncorrelated [16,31]. Some reports have also suggested that the type of reorientations in Phase II appears to be an intermediate admixture of Phases I and III [32-35]. When a $CH_3NH_3^+$ ion reorients by discrete angles of 120° about its C–N bond, its six protons trace out the equivalent of two complete current loops so an orbital magnetic moment $\mu_P$ along the C–N bond is generated depending on whether the reorientation is clockwise or counter-clockwise (pink arrows in Fig. 1). If adjacent $CH_3NH_3^+$ have $\mu_P$ pointing along the same directions, then they are bound to interact and contribute to the magnetic susceptibility $\chi(T)$. With decreasing temperature as the $C_4$ reorientations become forbidden, the probability of adjacent $\mu_P$ to point along the same directions increases. It is therefore of great interest to investigate $\chi(T)$ of MAPI and MAPB to learn about the degree of $\mu_P$ ordering, *viz.* the extent of proton orbital-orbital magnetic interactions in each of their three phases which in turn will allow us to obtain a better interpretation of the $CH_3NH_3^+$ dynamics in MAPI and MAPB. As will be shown below, magnetism based on $\mu_P$ interactions is identified to preside above room temperature in MAPI and in all of the polymorphs of MAPB and is believed to be the origin of their captivating nonlinear optical properties.

The syntheses of CH$_3$NH$_3$PbI$_3$ and CH$_3$NH$_3$PbBr$_3$ were grown according to existing literature [36] with slightly modified conditions:

Synthesis of CH$_3$NH$_3$PbI$_3$: A mixture of PbI$_2$ (4.61 g, 10 mmol) and CH$_3$NH$_3$I (1.59 g, 10 mmol) in 10 mL of γ-butyrolactone (GBA) was placed in a 20 ml vial, heated at 60 °C for 1 h, and then after filtration, the solution was heated at 110 °C for 5 h. The crystals were grown during this heating period. After cooled back to room temperature, the black-colored crystals were isolated by filtration.

Synthesis of CH$_3$NH$_3$PbBr$_3$: A mixture of PbBr$_2$ (3.67 g, 10 mmol) and CH$_3$NH$_3$Br (1.12 g, 10 mmol) in 10 mL of N, N-dimethylformamide (DMF) was placed in a 20 ml vial, heated at room temperature for 1 h followed by filtration. The solution was then heated at 80 °C for 16 h. After cooling down to room temperature, the orange-colored crystals were isolated by filtration.

The magnetic susceptibility of MAPI and MAPB were measured in the dark by the VSM option of a PPMS (Physical Property Measurement System) unit fabricated by Quantum Design, Inc. (U.S.A.). Samples 6-9 mg in weight were attached onto the standard quartz paddles by GE Varnish. The typical signals obtained for the magnetic moment were over $10^{-5}$ emu while the sensitivity of the equipment is $10^{-7}$ emu. The samples were always cooled and warmed at 1 K/min.

Figure 2a shows the temperature dependence of the molar magnetic susceptibility $\chi(T)$ of a MAPI single crystal under an external magnetic field of $H$ = 50 kOe along the $c$-axis direction. Two cooling and warming cycles in series are displayed to show its repeatability. The Supplementary Material file shows nearly exact results but with $H$ = 10–90 kOe. The most pronounced feature is the presence of a peak during warming with a maximum near 312 K followed by a drop of around 15%. The point of steepest descent was $T_{\text{II-I\_MAPI}}$ = 323.3 K which can only represent the phase transition from Phase II to I. The volume of the unit cell only changes by ~0.08% at $T_{\text{II-I\_MAPI}}$ [37] so the anomalous change in $\chi(T)$ at $T_{\text{II-I\_MAPI}}$ can only be interpreted as a direct reflection of the magnetic order-disorder process of the proton orbitals. This is because such anomalous behavior cannot stem from conventional antiferromagnetic ordering because even trace amounts of impurities would render $\chi(T)$ to become positive. In the Supplementary Material file we also show experimental data of the complex dielectric constant of a different MAPI crystal from the same batch which clearly shows that the structural phase transitions between I and II occurred at $T_{\text{I-II\_MAPI}}$ = 327.8 K and $T_{\text{II-I\_MAPI}}$ = 333.2 K. However, there was no coinciding discontinuity in $\chi(T)$ at $T_{\text{I-II\_MAPI}}$ during cooling and the peak was suppressed and shifted to around 286 K so in terms of degree of magnetic ordering, the $CH_3NH_3^+$ appeared to remain largely disordered when cooled back to its tetragonal phase near room temperature from $T_{\text{I-II\_MAPI}}$. In contrast, freshly synthesized samples possessed a high degree of order (marked by an * in Fig. 2a). From such, we could assess that proton orbital-orbital interactions are present at room temperature in MAPI and the extent of which the $CH_3NH_3^+$ are ordered depends on the thermal history of the sample.

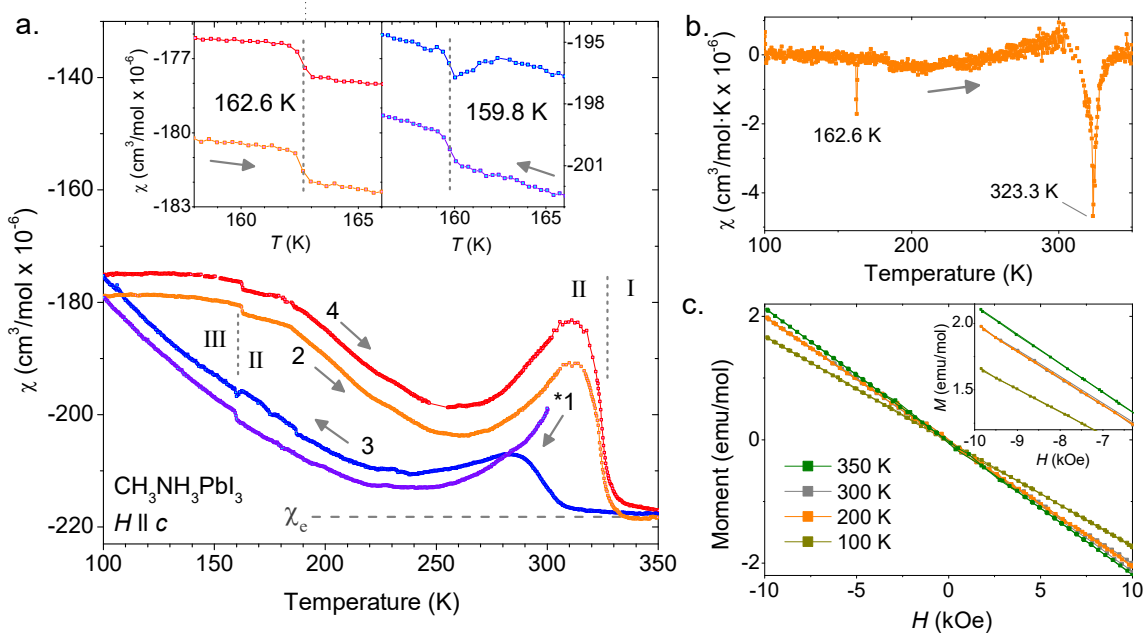

**Fig. 2.** (a) Molar magnetic susceptibility of $CH_3NH_3PbI_3$ under external magnetic field of $H$ = 50 kOe during cooling (curves 1 and 3) and warming (2 and 4) along the *c*-axis direction. An * marks the start of the scans. Vertical dashed lines are phase boundary lines. $\chi_e$ indicates the diamagnetic contribution of the electrons. Insets are zoomed-in views of the phase transition between phases II and III. (b) First order derivative of $\chi(T)$ (c) Magnetization vs. $H$ scans.

The phase transition between the tetragonal and orthorhombic phases manifested as a step-like discontinuity during cooling at $T_{\text{II-III\_MAPI}}$ = 159.8 K and warming at $T_{\text{III-II\_MAPI}}$ = 162.6 K (insets of Fig. 2a show enlarged views). The first order derivative of $\chi(T)$ is plotted in Fig. 2b to illustrate more clearly the first order nature of the two structural phase transitions. Figure 2c shows the molar magnetization $M$ vs. $H$ at 350, 300, 200 K and 100 K. The inset of Fig. 2c is an enlargement of the isothermals near 10 kOe. Reports on polycrystalline samples have shown that Pb site defects render the system paramagnetic [38] and even ferromagnetic [39]. A small amount of less than 0.1% of impurities usually completely masks the diamagnetic signal of the paired electrons $\chi_e$ (horizontal line in Fig. 2a) and that of the protons. In the present sample, $\chi_e$ was within 4.5% of the expected value of 230.59 x 10$^{-6}$ cm$^3$/mol, details of its calculation are available in the supplementary file. The negative slopes in Fig. 2c confirm our samples are largely diamagnetic due to the background contribution of the paired electrons so that the observed anomalies in $\chi(T)$ are due to proton orbital interactions. The exact contribution of the proton orbital interactions to the susceptibility can be obtained from $\chi(T) - \chi_e$.

Figure 3a displays the cooling and warming curves of $\chi(T)$ along the *b*-axis of an MAPB crystal under 50 kOe. The most distinguished feature is the step-up (step-down) discontinuity during warming (cooling) occurring exactly at the phase transition between the cubic and tetragonal structures at $T_{\text{II-I\_MAPB}}$ = 234.9 K ($T_{\text{I-II\_MAPB}}$ = 232.8 K). The first order derivative is plotted in Fig. 3b to show again the first order nature of the phase transitions. The volume of MAPB contracts by ~0.15% at $T_{\text{I-II\_MAPB}}$ [31] so the magnitude of $\chi(T)$ should increase by the same percentage; instead, the change is by ~2.5% so the step anomalies are undoubtedly signatures of the order-disorder transition of the proton orbitals.

The inset of Fig. 3a displays an enlargement of $\chi(T)$ near the phase transition between the orthorhombic and tetragonal phases. Small step-anomalies appeared at $T_{\text{II'-III\_MAPB}}$ = 146.8 K during cooling and $T_{\text{III-II'\_MAPB}}$ = 149.9 K during warming. In contrast, no apparent discontinuities were observed at the phase transition between the two tetragonal phases II and II'. Figure 3c shows *M*(*H*) scans at 100, 200 and 300 K; the magnitudes of the slopes are in good agreement with the $\chi(T)$ scans.

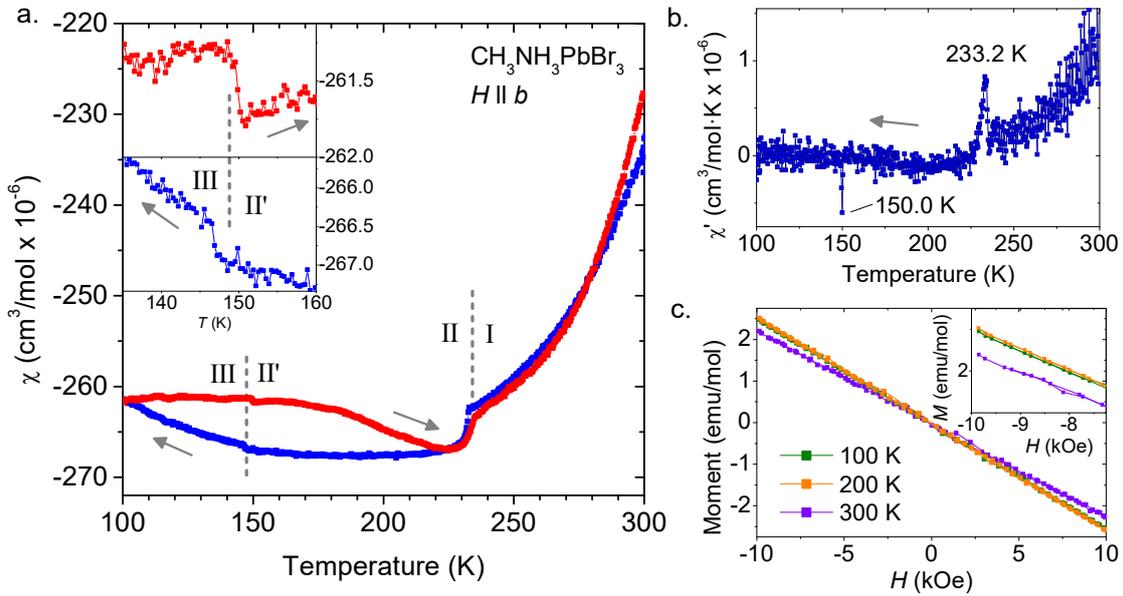

**Fig. 3.** (a) $\chi(T)$ of CH$_3$NH$_3$PbBr$_3$ along the *b*-axis direction under $H$ = 50 kOe during cooling and warming (indicated by arrows). Inset is an enlargement of the phase transition anomalies between phases II' and III. (b) First order derivative of $\chi(T)$. (c) Magnetization vs. *H* scans.

Akin to MAPI, the warming curve of $\chi(T)$ of MAPB did not retrace the cooling curve in Phases III and II suggesting a presence of glassy behavior [40]. The energies of the $C_3$ orbitals are very close to each other so the subsystems (each CH$_3$NH$_3^+$ molecule) in phases II and III cannot easily reach equilibrium causing a fraction of the subsystems to remain in a metastable state. Since the warming and cooling curves of the dielectric constant do overlap each other in phases II and III (Fig. S2), the breaking of ergodicity of the CH$_3$NH$_3^+$ appears to be magnetic in nature.

According to *ab initio* calculations and neutron diffraction measurements, the N protons form bonds with the Br ions in the cubic phase so the $CH_3NH_3^+$ primarily reorient along the (1,1,0) directions in Phase I [31, 41]. If this is the case, then MAPB should also exhibit magnetic anisotropy in the cubic phase. Figure 4a shows $\chi(T)$ of the same MAPB sample but measured along the *c*-axis direction; the general trends of the cooling and warming curves were fairly similar to the case for when $\chi(T)$ was measured along the *b*-axis (Fig. 3a) with the major exception that the magnitude of $\chi(T)$ for the latter case was on average 35% smaller. The expected diamagnetic contribution of the electrons $\chi_e$ of 182.59 x $10^{-6}$ $cm^3$/emu is also labelled in Fig. 4a to show that the *c*-axis $\chi(T)$ is less anomalous. The much larger deviation of up to 35% for the *b*-axis $\chi(T)$ appears to be because the $CH_3$ rotors are mostly confined to rotating freely in the *ab*-plane, so according to Lenz's law, the protons will rotate as to generate a magnetic field opposite of the applied field; the difference between the protons to electrons is that the protons will rotate in the opposite direction due to the polarity of the charge. An analogous effect also occurs in electrons circulating in benzene rings which enhances the magnitude of the magnetic susceptibility, often referred to as "exalted diamagnetism" [42]. As the temperature decreases from 300 K to just above $T_{I-II\_MAPB}$, the thermal fluctuations of the $CH_3$ rotors decrease which refines the orbitals of the protons allowing them to contribute more to the magnetic susceptibility with cooling.

As a comparison, the magnitude of $\chi(T)$ of MAPI along the *b*- and *c*-axes was nearly the same and the cubic phase was isotropic (Fig. S1d) apparently because the unit cell volume is large enough to allow the $CH_3NH_3^+$ to exhibit $C_3$ and $C_4$ reorientations as a whole. Returning to MAPB, several other distinguishing features were also found that remained fairly unchanged even under $H$ of up to 90 kOe (Fig. 4b): a) the structural phase transitions between the low-temperature tetragonal and orthorhombic phases were no longer observed; b) a cusp-like feature appeared around 121 K which coincides to the narrowing of the band gap occurring between 120 and 145 K [31]; and c) a change-in-slope discontinuity was observed only during warming near 183 K. These results lead us to conclude that significant magnetic interactions dependent on direction are present at all temperatures in MAPB even in the cubic phase above room temperature.

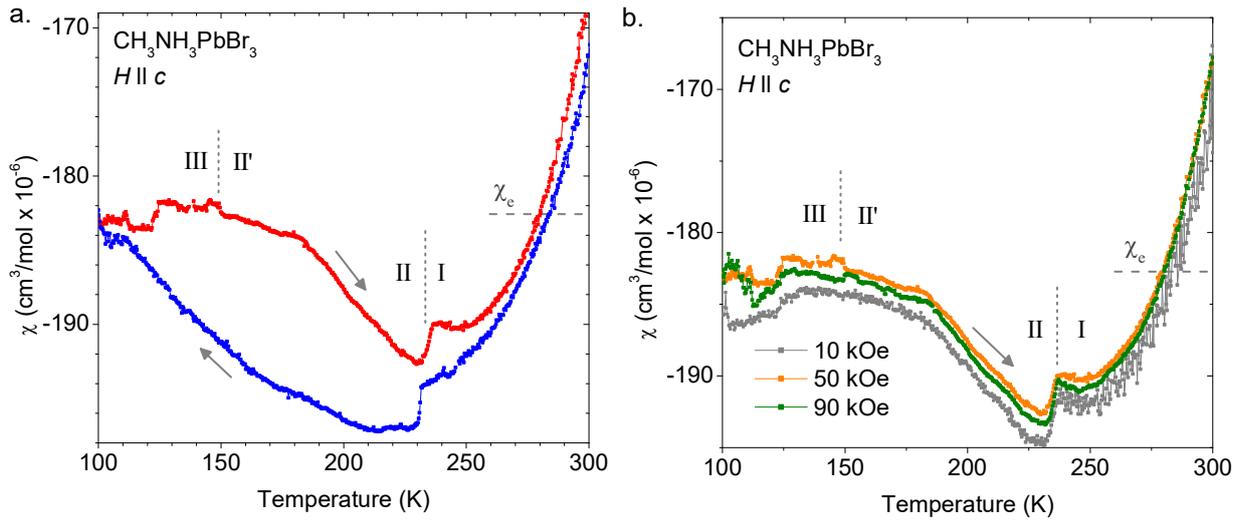

**Fig. 4.** (a) $\chi(T)$ of CH$_3$NH$_3$PbBr$_3$ along the $c$-axis direction under $H$ = 50 kOe during cooling and warming. (b) $\chi(T)$ under different $H$ during warming only.

The types of magnetic anomalies such as the sharp drop in $\chi(T)$ at the tetragonal to cubic phase transition in MAPI are unique in that no other type of system exhibits this sort of behavior. The magnitude of $\mu_p$ is weak, around 918 times smaller than that of the electron ($m_p/2m_e$, where $m_p$ and $m_e$ are the masses of the proton and electron, and the factor of 2 represents the equivalent of two current loops completed by six protons reorienting by 120°) but neighboring $\mu_p$ interact with each other because they are forcibly aligned by the crystal field. From such, the magnetization does not diverge at the order-disorder transitions as dictated by Landau theory because it is not the order parameter. The behavior of $\chi(T)$ is therefore more or less a reflection of the proton dynamics of the lattice and independent of $H$ as witnessed in Fig. 4b. Nevertheless, our results offer a possible explanation to the origins of the observed photoinduced magnetization in single crystals of MAPB [43] and thin films of MAPI [44, 45] in that the proton orbital-orbital interactions intensify (due to higher dynamics) when the highly light-absorbing $CH_3NH_3^+$ ions [28] are exposed to light. A direct implication of this is that an internal magnetic field $B$ can be established by the photoinduced magnetization $M$ even when $H = 0$ according to $B = H + 4\pi M$. According to the charge carriers of the inorganic framework moving at velocities $v$, $B$ is an applied field originating from the light-absorbing $CH_3NH_3^+$ ions so unconventional electromagnetic forces between the two interpenetrating frameworks appear to be the mechanism underlying the unique properties of long diffusion lengths [3, 4], high mobility [6] and reduced electron/hole recombination rates [7] in MAPI and MAPB.

To conclude, we identified the presence of magnetic interactions of the $CH_3$–$CH_3$ and $NH_3^+$–$NH_3^+$ type in all of the polymorphs of $CH_3NH_3PbI_3$ and $CH_3NH_3PbBr_3$ single crystals from magnetic susceptibility measurements. In the ordered phases of both systems, the degree of $CH_3NH_3^+$ ordering is highly dependent on thermal history. For the case of $CH_3NH_3PbBr_3$, there exists magnetic anisotropy even in the cubic phase apparently due to the $CH_3$ and $NH_3^+$ units reorienting in an uncorrelated manner. The simultaneous presence of magnetic ordering of the proton orbitals and onset of ferroelectricity classifies MAPB as a single-phased multiferroic with an exotic type of magnetoelectric coupling effect that is not based on electron spin. Identification of proton orbital-orbital interactions in MAPI and MAPB provides us with a missing link in that the exceptional properties of hybrid organic-inorganic perovskites are likely 'opto-magneto-electronic' in nature.

**Declaration of Competing Interest**

The authors declare no competing financial interests.

**Acknowledgements**

This project was partially funded by a) a General Project grant of the Shenzhen Universities Sustained Support Program No. GXWD20201230155427003-20200822225417001 from the Science, Technology and Innovation Commission of Shenzhen Municipality and b) a National Natural Science Foundation of China grant No. 21702038.

**Supplementary material**

# Magnetic interactions based on proton orbital motion in $CH_3NH_3PbI_3$ and $CH_3NH_3PbBr_3$

Lei Meng,[a,†] Miao Miao Zhao,[a,†] Yi Yang Xu,[a,†] Chu Xin Peng,[a,†] Yang Yang,[a] Tian Tian Xing,[a] Peng Ren,[a,*] and Fei Yen[a,*]

a. Department State Key Laboratory on Tunable Laser Technology, Ministry of Industry and Information Technology Key Laboratory of Micro-Nano Optoelectronic Information System and the School of Science, Harbin Institute of Technology, Shenzhen. University Town, Shenzhen, Guangdong 518055, P. R. China.

# 1. Additional magnetic susceptibility data of CH$_3$NH$_3$PbI$_3$ along the *b* and *c*-axes

Figure S1a-S1c show the molar magnetic susceptibility $\chi(T)$ of CH$_3$NH$_3$PbI$_3$ (MAPI) under external magnetic fields of $H$ = 10, 50 and 90 kOe along the *c*-axis in the temperature range of 5 – 350 K. This sample was different from that of Figure 2a but both crystals were from the same batch. The critical temperatures and shapes of the phase transitions between the cubic (I) and tetragonal (II) as well as tetragonal and orthorhombic (III) of the two crystals were nearly identical. The experimental methods and conditions were also the same. The main difference between the measurements of Figure S1a-S1c and Figure 2a is that in the former case the crystal was cooled from 300 K down to 5 K and back to 300 K; in contrast to the latter case, the sample was cooled from 300 K but only down to 100 K and then back to 300 K.

Figure S1d displays $\chi(T)$ along the *b*-axis and *c*-axis (curves 1 and 3 from Fig. 2a) of CH$_3$NH$_3$PbI$_3$. The discontinuities accompanying the phase transitions were rather similar in both cases. However, the peak anomaly with a maximum near 312 K became less pronounced while a shoulder-like feature developed around 287 K. Lastly, the cubic phase was nearly isotropic in contrast to the case of CH$_3$NH$_3$PbBr$_3$: the reason behind this discrepancy is due to the CH$_3$NH$_3^+$ reorienting more or less correlated and uncorrelated in CH$_3$NH$_3$PbI$_3$ and CH$_3$NH$_3$PbBr$_3$, respectively.

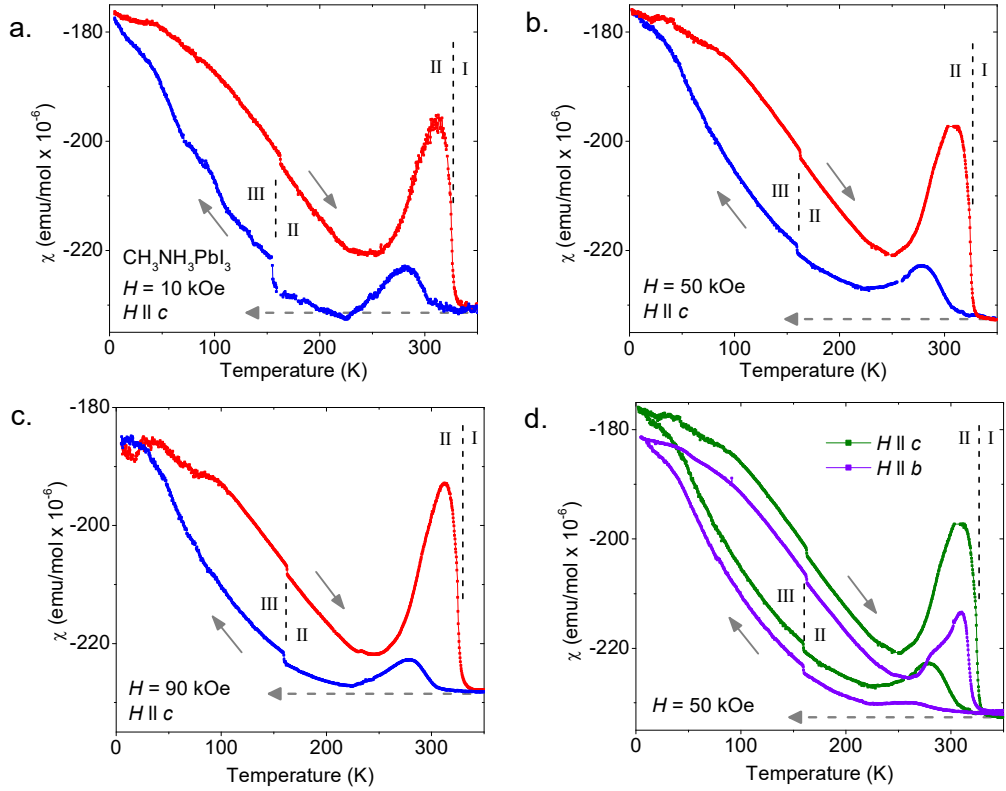

**Fig. S1**. Molar magnetic susceptibility $\chi(T)$ of $CH_3NH_3PbI_3$ with respect to temperature under external magnetic fields of (a) $H$ = 10 kOe, (b) 50 kOe and (c) 90 kOe along the *c*-axis. The cubic, tetragonal and orthorhombic phases are labelled as I, II and III, respectively. Vertical dashed lines demarcate the phase boundaries. Arrows show the cooling and warming curves. (d) $\chi(T)$ along the *b*-axis and *c*-axis under $H$ = 50 kOe. Horizontal dashed lines are expected diamagnetic susceptibility values obtained from Pascal constants to within 5%.

## 2. Determination of the expected diamagnetic contribution of electrons to the magnetic susceptibility

The diamagnetic contribution from the electrons to the magnetic susceptibility can be obtained by adding the expected contributions of each atom and bond of the system, also known as Pascal's constants. Table S1 details the contribution of each of the components for $CH_3NH_3PbI_3$ and $CH_3NH_3PbBr_3$ for an expected magnetic susceptibility of 230.6 x 10⁻ $6$ and 182.6 x 10$^{-6}$ emu/mol, respectively. This value remains independent of temperature provided thermal expansion effects are negligeable so the contributions from molecular dynamics can be obtained by subtracting this base value from the measured magnetic susceptibility.

|  | $\chi_{MAPI}$ | $\chi_{MAPBr}$ |
|---|---|---|
|  | X 10$^{-6}$ emu/mol | X 10$^{-6}$ emu/mol |
| H atoms (x3) | –2.93 x 3 = –8.79 | –2.93 x 3 = –8.79 |
| C atom | –6.00 | –6.00 |
| C – N bond | 0.00 | 0.00 |
| NH$_3$ | –18.00 | –18.00 |
| Pb atom | –46.00 | –46.00 |
| I atom (x3) | –50.60 x 3 = –151.80 |  |
| Br atom (x3) |  | –34.6 x 3 = –103.8 |
| Total | 230.59 | 182.59 |

**Table S1:** Diamagnetic contributions of the atoms and bonds present in CH$_3$NH$_3$PbI$_3$ $\chi_{MAPI}$ and CH$_3$NH$_3$PbBr$_3$ $\chi_{MAPBr}$.

### 3. Complex dielectric constant measurements of CH$_3$NH$_3$PbI$_3$

Figures S2a and S2b show the near-static complex dielectric constant of MAPI along the $b$-axis at 1 kHz. A sharp drop occurs in the real part $\varepsilon'(T)$ by nearly three-fold during cooling at the tetragonal to orthorhombic phase transition $T_{II-III\_MAPI}$ = 158.4 K. During warming, the same phase transition occurred at $T_{III-II\_MAPI}$ = 163. 4 K. These results are in good agreement with those of existing literature [25]. The imaginary part $\varepsilon''(T)$ exhibited much less pronounced discontinuities shown in the inset of Figure S2b. At the phase transition between the cubic and tetragonal phases, small step discontinuities were observed at $T_{\text{I-II\_MAPI}}$ = 327.8 K and $T_{\text{II-I\_MAPI}}$ = 333.2 K and during cooling and warming, respectively, in both $\varepsilon'(T)$ and $\varepsilon''(T)$ (Figures S2c and S2d). We note that $T_{\text{I-II\_MAPI}}$ and $T_{\text{II-I\_MAPI}}$ have never been observed in existing literature.

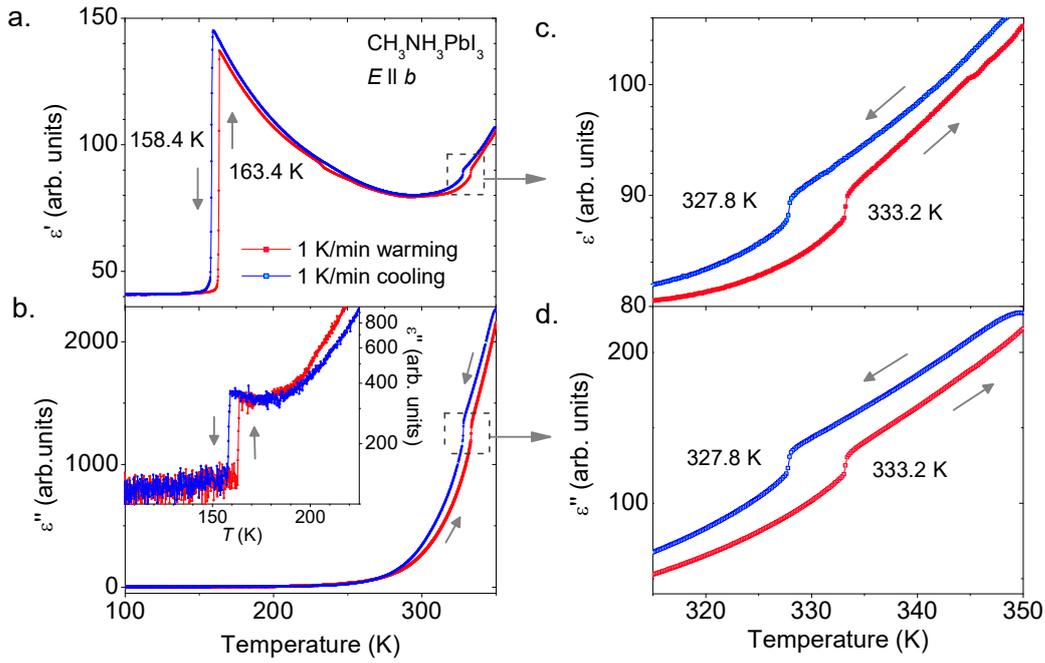

**Fig. S2**. a) Real $\varepsilon'(T)$ and b) imaginary $\varepsilon''(T)$ parts of the dielectric constant at 1 kHz along the b-axis direction. Enlarged regions of $\varepsilon'(T)$ and $\varepsilon''(T)$ near the phase transition between the cubic and tetragonal phases are shown in c) and d).

## 4. Crystal structure of CH$_3$NH$_3$PbI$_3$ and CH$_3$NH$_3$PbBr$_3$ at 110 and 300 K

To ensure the quality of the single crystals, XRD measurements were performed on the samples at room temperature and at 110 K; the lattice constants are provided in Table S2. At 110 K, the structure of both systems is orthorhombic coinciding to their respective phases III. At 300 K, CH$_3$NH$_3$PbI$_3$ is tetragonal and CH$_3$NH$_3$PbBr$_3$ is cubic, both of which coincide to their respective phases II and I.

|  |  | a (Å) | b (Å) | c (Å) | α (°) | β (°) | γ (°) |
| --- | --- | --- | --- | --- | --- | --- | --- |
| CH$_3$NH$_3$PbI$_3$ | 300 K | 8.87731 | 8.87731 | 12.66766 | 90 | 90 | 90 |
|  | 110 K | 8.86352 | 12.61456 | 8.60448 | 90 | 90 | 90 |
| CH$_3$NH$_3$PbBr$_3$ | 300 K | 5.93310 | 5.93310 | 5.93310 | 90 | 90 | 90 |
|  | 110 K | 7.99289 | 11.85312 | 8.58281 | 90 | 90 | 90 |

**Table S2**. Lattice parameters and space groups of CH$_3$NH$_3$PbI$_3$ and CH$_3$NH$_3$PbBr$_3$ at 300 and 110 K.